\documentclass[11pt,leqno]{article}
\usepackage{amssymb}
\usepackage{amsmath}
\usepackage[numbers]{natbib}\def\citet{\cite}
\def\xxxonly{\comm}
\def\xxxonly{ }

\newcommand{\comm}[1]{}

   \csname @addtoreset\endcsname{equation}{section}
    \csname @addtoreset\endcsname{figure}{section}
    \csname @addtoreset\endcsname{table}{section}

\newcounter{thanksnum}
\def\thanksnumber#1
{\setcounter{thanksnum}{\value{footnote}}\setcounter{footnote}{#1}%
                     \addtocounter{footnote}{-1}\footnotemark
                     \setcounter{footnote}{\value{thanksnum}}}

\newtheorem{theorem}{Theorem}[section]
\newtheorem{lemma}{Lemma}[section]

\newtheorem{proposition}{Proposition}[section]
\newtheorem{corollary}{Corollary}[section]

\newtheorem{remark}{Remark}[section]

\newtheorem{example}{Example}[section]

\def\e{\varepsilon}

\def\defi{\stackrel{{\scriptscriptstyle \Delta}}{=}}

\def\a{\alpha}

\def\o{\omega}
\def\O{\Omega}

\def\F{{\cal F}}
\def\w{\widehat}
\def\Ind{{\,\rm Ind\,}}
\def\Ind{{\mathbb{I}}}

\def\sign{{\rm  sign\,}}

\def\esssup{\mathop{\rm ess\, sup}}
\def\const{{\rm const\,}}

\def\R{{\bf R}}
\def\E{{\bf E}}
\def\P{{\bf P}}

\def\H{{\cal H}}

\def\L{L}

\def\b{\beta}

\def\g{\gamma}

\def\ww{\widetilde}
\def\X{{\cal X}}
\def\t{\theta}
\def\oo{\bar}

\def\D{{\Delta}}
\def\p{\partial}
\def\G{\Gamma}

\def\U{{\cal U}}
\def\V{{\cal V}}

\def\L{{\cal L}}

\newcommand{\be}{\begin{equation}}
\newcommand{\ee}{\end{equation}}
\newcommand{\bd}{\begin{displaymath}}
\newcommand{\ed}{\end{displaymath}}
\newcommand{\ba}{\begin{array}{ll}}
\newcommand{\ea}{\end{array}}
\newcommand{\baa}{\begin{eqnarray}}
\newcommand{\eaa}{\end{eqnarray}}
\newcommand{\baaa}{\begin{eqnarray*}}
\newcommand{\eaaa}{\end{eqnarray*}}

\def\oo{\bar}

\def\wD{{\w D}}

\def\U{{\cal U}}
\date{\ Submitted: March 22, 2016; revised: October 25, 2018}
\title{First Order BSPDEs in higher dimension for optimal control problems}
\author{
Nikolai Dokuchaev\footnote{
School of Electrical Engineering, Computing and Mathematical Sciences, Curtin
University,  GPO Box U1987, Perth, 6845 Western Australia,  email
N.Dokuchaev@curtin.edu.au}
}
\begin{document}
\maketitle
\let\thefootnote\relax\footnote{
This is an author-produced  extended version  of an article published in
SIAM Journal on Control and Optimization (2018) {\bf 55} (2), 818-834. In this version, Section 6 has been extended. }
\begin{abstract} The paper studies the First Order BSPDEs (Backward Stochastic Partial Differential Equations)
suggested earlier for a case of multidimensional state domain with a boundary.
These equations represent analogs of Hamilton-Jacobi-Bellman  equations
and allow to construct the value function for stochastic optimal
control problems  with unspecified dynamics where the underlying processes
do not necessarily satisfy  stochastic differential equations of a known kind with a given structure. The problems considered
arise in financial modelling.

\end{abstract}
%
\section{Introduction}
In Bender and Dokuchaev (2016a,b), \index{\citet{BD1,BD2}} some special
first order backward stochastic partial differential equations (BSPDEs)
were suggested and studied for the case of one dimensional state variable.
These equations were used for optimal value functions for pricing problems for swing options with
very mild assumptions on the underlying payoff process. The present paper extends these results on multidimensional case.

Stochastic Partial Differential Equations (SPDEs)
 are well studied in the literature, including
 the case
 of  forward and backward  equations; see, e.g.,
\index{cite{Alos,Ba,DaPT,CK,CM,CMG, D92,D95,D05,D08b,D10,D12,DT, Duan,FZ, A1,Hu,kr,Ma,Ma99,Mas,Par,Roz,Wa,Zh}
 and the references
therein.} Walsh (1986)
Al\'os et al
(1999),
 Chojnowska-Michalik (1987),  Rozovsky (1990),  Zhou (1992),  Pardoux (1993), Bally {\it et al} (1994),  Chojnowska-Michalik and Goldys
 (1995),  Maslowski (1995), Da Prato and Tubaro (1996),
Gy\"ongy (1998), Krylov (1999),   Mattingly (1999),
  Duan {\em et al} (2003),  Caraballo {\em et al } (2004), Dokuchaev (2005), Mohammed  {\em et al} (2008), Feng and Zhao (2012),
and the bibliography therein.

Backward Stochastic Partial Differential Equations (BSPDEs) represent versions of
the so-called Bismut-Peng  equations where the diffusion
term is not given a priori but needs to be found; see e.g.   Hu and Peng (1991),  Peng
(1992),  Zhou (1992),  Dokuchaev (1992,1995,2008,2010, 2011,2012,2015a,b), Du  and Tang (2012), Du {\em at al} (2013), Hu {\it et al} (2002),  Ma and
Yong (1999),  and the bibliography therein.  \index{The existing literature considers  BSPDEs of the
second order (and their generalizations) such that (i) the matrix of
the higher order coefficients is  positive, and (ii) the so-called
coercivity condition holds. }

Some additional conditions on
the coercivity are usually imposed in the literature; see e.g.
condition (0.4) in Rozovsky (1990),  Ch. 4. Without these
conditions, a parabolic type SPDE is regarded as degenerate. These  degenerate equations of the second order (with a relaxed
coercivity condition) were also widely studied; see the
bibliography in Du {\em et al} (2013), Du and Zhang (2013), Dokuchaev (2015a). For the
degenerate backward SPDEs in the whole space, i.e., without
boundaries, regularity results were obtained in   Ma and
Yong (1999), Hu {\it et al} (2002),  Du  and Tang (2012), Du {\em at al} (2013).
\index{
  citet{Roz},
citet{MaYong}, citet{Hu}, and more recently by citet{DTZ} and
citet{DZ}.}
Some special
first order forward SPDEs without boundary  were  considered in Gikhman and Mestechkina (1983) and Kunita (1990) by the method of characteristics, and  in Hamza and Klebaner (2006). The methods developed in these works cannot be applied
in the case of a domain with boundary  because of regularity issues
that prevent using an approximation of the differential operator  by
a non-degenerate one. It turns out that the theory of degenerate
SPDEs in domains is much harder than in the whole space and was, to
the best of our knowledge, not addressed yet in the existing literature, except for first attempts in  Bender and Dokuchaev (2016a,b) and in Dokuchaev (2015a).
The present paper also considers  a problem of
this kind.

It is common to use BSPDEs  as stochastic analogs of the backward Kolmogorov-Fokker-Planck
equations or related Hamilton-Jacobi-Bellman  equations known for controlled Markov diffusion type processes.
In non-Markovian  control problems,  the backward Hamilton-Jacobi-Bellman  equations
equations have to be replaced by corresponding backward SPDEs; this was first
observed  by Peng (1992)
\index{Pe1992} in a setting with
backward SPDEs of the second order
equations such that the matrix of the higher order coefficients is
positive definite.

In Bender and Dokuchaev
(2016a,b), new special  First order BSPDEs were introduced. They represented  analogs of Hamilton-Jacobi-Bellman  equations for some non-Markovian
stochastic
optimal control problems associated with pricing of swing options in continuous time.
These equations are not exactly differential, since their solutions can be discontinuous in time, and they allow
 very mild conditions on the
underlying driving stochastic processes with unspecified dynamics.
 More precisely, the method does not have to assume a particular
evolution law of the underlying process;   the underlying processes
do not necessarily satisfy  stochastic differential equations of a known kind with a given structure.
In particular, the  First Order BSPDEs describe the value function
even in the situation where the underlying price process cannot be
described via a stochastic equation ever described in the
literature. The numerical solution requires just to calculate certain conditional expectations
of the functions of the process without using its evolution law (see the discussion in Section 4).
It can be also noted that these equations are not the same as the first order deterministic HJB equations known in the deterministic optimal
control.

In the present paper, we  extend
the approach suggested in Bender and Dokuchaev
(2016a,b) and derive some First order BSPDEs for multi-dimensional domains using a different  proof.
Again, the suggested BSPDEs  represent analogs of HJB  equations for some non-Markovian
stochastic
optimal control problems associated with  applications in financial modelling.
The paper establishes existence of solutions for these equations and the fact the value functions for underlying control problems
satisfy these equations. Some numerical methods are discussed.

\section{Problem setting}
Consider a probability space $(\O,\F,\P)$, $\O=\{\o\}$.
 Let  $X(t)=(X_1(t),...,X_n(t))^\top$ be a current random process with the values in $\R^n$ such that $X(t)$ is RCLL (right continuous with left limits) process,   $\E
\sup_{t\in[0,T]}|X(t)|^2<+\infty$, and $X_i(t)\ge 0$ a.s. for all $t$ and $i$. \index{we assume that it is a
martingale.}

 Let $\{\F_t\}_{t\ge 0}$ be the filtration generated by $X(t)$.

We emphasize that an evolution equation for $X$ is not specified, similarly to the setting from Bender and Dokuchaev (2016a,b). For instance, we do not assume that $X(t)$ is solution of an It\^o's equation with particular drift and diffusion coefficient, or
of any other equation such as jump-diffusion equation,  etc.  The case where dynamics of $X$ is described by one of these equations is not excluded; however, one does  not need the structure and  parameters
of these equations for our analysis. This is an unusual setting  for stochastic control and theory of HJB equations.
As is discussed in Section \ref{secN} below, one has  to know conditional distributions of $X(t_{k+1})$ given $\F_{t_k}$ and a sampling sequence
$\{t_k\}\subset[0,T]$ for numerical implementation of the analytical results.

Let a positive integer $n$ be given.

For $x=(x_1,...,x_n)^\top\in\R^n$ and $y=(y_1,...,y_n)^\top\in\R^n$, we write $x\preceq y$ if and only if $x_i\le y_i$ for all $i$.

Let $\G_0\subset \R^n$ be a closed convex conic set such that if $x\in\R^n$, $x\preceq \w x$ and $\w x\in \G_0$ then $x\in \G_0$, and such that,
for any  $x=(x_1,...,x_n)^\top\in\G_0$ and any $j\in\{1,...,n\}$, there exists $M>0$ such that  $x=(x_1,...,x_j+M,...,x_n)^\top\notin\G_0$.
Let $\w g\in \R^n$ be a given vector with positive components, and let $\G=\{y\in\R^n:\ y=x+\w g, \ x\in \G_0\}$.

\begin{example} In particular, the following choices of $\G$ are admissible.
\begin{enumerate}
\item  $\G=\{x=(x_1,...,x_n)^\top\in\R^n:\ x_i\le \a_i, \ i=1,...,n\}$, where $\a_i>0$ are given.
\item $\G=\{x\in\R^n:\ x^\top a\le 1\}$, where $a\in\R^n$ is a given vectors with positive components.
\item  $\G=\{x=(x_1,...,x_n)^\top\in\R^n:\ x^\top a_j\le 1, \ j=1,...,m\}$, where $m>0$, and
where $a_j\in\R^n$ are given non-zero vectors with non-negative components,  such that, for any $i\in\{1,...,n\}$,
there exists $j$ such that $i$th component of $a_j$ is positive.
\end{enumerate}
\end{example}

Let  $K\subset\R^n$ be a given convex \index{compact} set.

For $y\in\G$, let $U(t,y)$ be the set of processes $u(s)=(u_1(t),...,u_n(t)):[t,T]\times \O\to K$
being adapted to $\F_t$ and such that $y+\int_t^T  u(s)ds\in \G$ a.s.

 Let a function $f(x,u,t):\R^n\times K\times[0,T]\to \R$ be given.

 We consider two cases:
 \begin{enumerate}
 \item $f(x,u,t)=u^\top x$ and $K=[0,L]^n$ for some given $L>0$. \index{$K$ is compact and
such that if $v=(v_1,...,v_n)^\top\in K$ then $v_i\ge 0$, $i=1,...,n$.}
\item $f(x,u,t)=u^\top x-u^\top G u$, \xxxonly{$m=n$,}
where $G$ is a symmetric
positive-definite matrix, and $K=\R^n$.
\end{enumerate}
These conditions seems very special, however, they cover many important optimization problems arising in mathematical finance.
The special  case where $G=0$ and $n=1$ covers the pricing problem for swing options considered in Bender and Dokuchaev (2016a,b).

\subsection{Optimal control  problem}
For given $y\in\G$ and $t<T$, we consider the problem
\baa
\hbox{Maximize}\quad \E F(u,t)\quad\hbox{over} \quad u\in U(y,t).
\label{problem}
\eaa
Here
 \baa
F(u,t)=\int_t^Tf(X(s),u(s),s)ds. \label{Foptg} \eaa
\index{\def\UU{V}
Let $\UU$ be a $C_1$-smooth non-decreasing concave function $U:\R\to\R$ bounded together with its derivative, and let. Consider a payoff \baa
F(u,t)=\UU\left(\int_t^Tu(s)^\top X(s)ds\right), \label{Foptg} \eaa
 where $X(t)$ is a current payoff process,
\baaa u\in \cup_{y\in[0,1]}\U(t,y).  \eaaa  The function $u(t)$ is
the control process.}
Let $y$ be a $\F_t$-measurable random variable with values in
$[0,1]$, and let \baa J(t,y)=\esssup_{u\in U(t,y)}\E_t F(u,t).
\label{J}\eaa In other words, this is the value function for problem (\ref{problem}). Here $\E_t=\E\{\cdot|\F_t\}$; we use this notation for brevity.
\subsubsection*{Applications in finance}

 In Bender and Dokuchaev (2016a,b), it was shown that the  above introduced
optimization problem (\ref{problem}) with  $n=1$, $K=[0,L]$, and $G=0$, gives a solution
for  pricing problem for swing options with underlying payoff process $X(t)$. In this setting, $u(t)$ is the exercise rate selected by the option holder.
The swing option holder wishes to maximize the expected cumulative payoff by selection the distribution  in time $u(t)$ of the exercise rights.
In addition, it was shown  that this problem
 can be used for approximation of prices for multi-dimensional American options, with the choice of
 $L\to +\infty$.
 Similarly,  problem (\ref{problem})
 with  $n>1$ can be interpreted as a pricing problem
for swing options on consumptions of $n$ different types of energy.
The choice of $G\neq 0$ can be used to  model the settings where there are no hard constraints
on the rates of exercises, with a penalty for excessive rates instead.
Again, these problems can be used for approximation of the classical solutions for the pricing problems  for multi-dimensional American options. For this, we have can consider
 either   $G\to 0$ and $K=\R^n$ or  $G\neq 0$, $K=[0,L]^n$, and $L\to +\infty$.

\index{With some extra modification,  multi-dimensional Asian options an d optimal portfolio selection problems aslo can be covered.}

Another possible setting is where the swing option holder has to maximize the expected cumulative payoff
but where there is a preferable vector of exercise  rates $\ww u(t)$; it could be a currently observable random process defined by dynamically changing
 technological or market conditions.
Let $X(t)$ be the current vector of the underlying payoffs.
The swing option holder wished to maximize
\baaa
\int_0^T[X(t)^\top u(t)-(u(t)-\ww u(t))^\top G (u(t)-\ww u(t))]dt.
\eaaa
The term $(u(t)-\ww u(t))^\top G (u(t)-\ww u(t)$ represents some penalty for mismatching the preferable rate $\ww u(t)$.
This is equivalent to the maximization of
 \baaa
\int_0^T[X(t)^\top u(t)+2\ww u(t)^\top G u(t)- u(t)^\top G u(t)]dt= \int_0^T[\ww X(t)^\top u(t)- u(t)^\top G u(t)]dt,
\eaaa
where $\ww X(t)=X(t)+2G\ww u(t)$.  This is a special case of the problem introduced above.

Some related setting for optimal energy trading was developed in Dokuchaev (2015c).

\index{ On the conditional probability space
$(\O,\P(\cdot|\F_t),\F)$, we consider  the Hilbert space $\H_t$ of
all square integrable and adapted processes from
$L_2([t,T]\times\O,\P(\cdot|\F_t),\F)$, with the "conditional" norm
$\|u\|_{\H_t}=\left(\E_t\int_t^T|u(s)|^2ds\right)^{1/2}$.
\par}
\subsection*{Existence of optimal $u$ and properties of $J$}
\begin{lemma}\label{lemmaE}
 For every pair $(t,Y)$, where $t$ is a stopping time and $Y$ is a $\mathcal{F}_t$-measurable random vector with  values in $\G$,
there is an optimal control $\bar u \in U(t,Y)$.
\end{lemma}

\par
Let $u^{t,y}(s)$ be an optimal control for (\ref{J}) (or one of the
optimal controls).

\begin{lemma}\label{lemmaJ}
\begin{itemize}
\item[(i)]  For  $t\in[0,T]$ and $\oo y\in\G$,   $\sup_{y\in \G:\ \oo y\preceq y,\,t\in[0,T]}\esssup_\o|J(t,y)|\le\const$.
\item[(ii)] For any $t\in[0,T]$ and $y,\w y\in\G$ such that $y\preceq \w y$, we have that $J(t,\w y)\le J(t,y)$ a.s.
\item[(iii)]  For any $t\in[0,T]$ and $j\in\{1,...,n\}$, $J(t,y)$ is almost surely Lipschitz in $y$ uniformly in any bounded subset of $[0,T]\times\G$.
\end{itemize}
\end{lemma}

\begin{lemma}\label{lemmaconc}  The function $J(t,y)$ is concave in $y\in\G$,
a.s. for all $t\in[0,T]$.
\end{lemma}
\par
By Lemma \ref{lemmaconc}, the left-hand partial derivatives $D^-_{y_j}J(t,y)$ and the right-hand partial derivatives $D^+_{y_j}J(t,y)$ exist and are
uniquely defined.
\index{\begin{lemma}  For non-random $(t,y)$, the function $D^-_y(t,y)=0$ if $1-y>L(T-t)$.
\end{lemma}
\par
{\em Proof.} Clearly, the optimal strategy is $u^{t,y}(s)=L$ for
a.e. $s\in[t,T]$ a.s. Then $J(t,y)=\E_t\int_t^TL X(s)ds$ does not
depend on $y$. $\Box$
\begin{lemma}
For non-random $(t,y)$,  $J'_y(t,y-0)=0$ and
$J'_y(t,y+0)=-\esssup_{s\in[t,T]}X(s)?????$ if $1-y=L(T-t)$.
\end{lemma}
\par
{\em Proof.}  Again, the optimal strategy is $u^{t,y}(s)=L$ for a.e.
$s\in[t,T]$ a.s. Then $J(t,y)=\E_t\int_t^TL
X(s)ds=X(t)L(T-t)=X(t)(1-y)$. $\Box$
\begin{corollary}
For non-random $(t,y)$, $J'_y(t,y)\le -\esssup_{s\in[t,T]}X(s)$ a.s.
for a.e. $t\in [1/L-T,T]$.
\end{corollary}}
\section{The main result}

\def\myD{\w D}
\def\myD{D^+}
 We denote by $D_y^\pm J(s,y)$ the vector columns $\{ D_{y_i}^\pm J(s,y)\}_{i=1}^m$ in $\R^n$. We denote $(x)_+=\max(x,0)$.

  Let $\p\G$ be the boundary of $\G$.
 \begin{theorem}\label{ThM}
The value function
$J$ satisfies following first order BSPDE in $(t,y)$:
\baa
&& J(t,y)=\E\left\{\left. \int_t^T \sup_{v\in K}(f(X(s),v,t)+ v^\top  \myD_y J(s,y))ds\right|\mathcal{F}_t\right\}, \qquad t<T,\quad y\in\G,\nonumber \\
&&J(t,y)|_{y\in\p\G}=0.
\label{BSPDE}\eaa
\end{theorem}

In particular, if  $f(t,x,u)=u^\top X$ and $K=[0,L]^n$, then the equation has the form
\baaa
&& J(t,y)=L\E\left\{\int_t^T \left(\sum_{i=1}^n(X_i(s)+  \myD_{y_i} J(s,y))_+\right)ds\Bigl|\mathcal{F}_t\right\},\nonumber\\ &&\qquad t<T,\quad y\in\G,\nonumber \\
&&J(t,y)|_{y\in\p\G}=0.
\label{BSPDE2}\eaaa
If  $K=\R^n$ and $f(t,x,u)=u^\top X-u^\top G u$, then the equation has the form
\baaa
&& J(t,y)= \frac{1}{4}\E\left\{\int_t^T(X(s)+ \myD_y J(s,y))^\top G^{-1}(X(s)+ \myD_y J(s,y))ds\Bigl|\mathcal{F}_t\right\},\nonumber\\ &&\qquad t<T,\quad y\in\G,\nonumber \\
&&J(t,y)|_{y\in\p\G}=0.
\label{BSPDEG}\eaaa
\begin{remark}
Theorem \ref{ThM} and Lemma \ref{lemmaE} imply existence of solution of problem (\ref{BSPDE}). However, it does not state its uniqueness.
For a special case of $n=1$ and $G=0$, the uniqueness was established in Bender and Dokuchaev (2016b); this task
required significant analytical efforts.  We leave the problem of uniqueness of the solution for $n>1$
for further research.
\end{remark}

\section{On numerical implementation}\label{secN}
For Markov models, corresponding HJB equations can be solved using finite difference. Unfortunately, this approach may not be effective for a large
number of parameters describing the dynamics of the Markov model.
The features of First Order BSPDEs (\ref{BSPDE}) allow to use some alternative methods described below.
\subsection*{On numerical feasibility for  an unspecified dynamics law for $X(t)$}
Equation (\ref{BSPDE})
can be solved after discretization backward in $t$ using the first order finite differences in $t$ and $y$, with
calculation of the conditional expectation on each step by the Monte-Carlo method.
This approach allows to use the following attractive feature of equation (\ref{BSPDE}): the solution of this  equation can be calulated
even for models where  the dynamics law for $X(t)$ is not specified.
For this,  one needs to know, for a time-discretization sequence $\{t_k\}$, the conditional distributions of $X(t_{k+1})$ given $\F_{t_k}$.
  \par
  The benefit is that we do not need a hypothesis about the dynamics of $X$, or the equation for its evolution, i.e. if it is an It\^o's equation, etc. Moreover, there are models where the information about the distribution  of  $X(t_{k+1})$
 is more accessible and reliable than the information  about the dynamics law.
{\em In general, the dynamics law for $X$ is more difficult to establish since it is not robust with respect to the
variations of the probability distribution as can be seen from the following example.} \par\noindent
\begin{example}
 Let $X$ be a Wiener process. This process
can be approximated  by  pathwise absolutely continuous  processes $X_\e(t)$ that will be, therefore, statistically
indistinguishable from $X$ and yet  will have  a very different dynamics law, without the amazing features of the It\^ o's processes.
\end{example}
\par
If the dynamics of $X(t)$ are described by a particular equation (such as an It\^o's equation), then
the parameters of the equation will define the resulting conditional expectation but does not have to be used directly.

\subsection*{Estimation of $J$ using pathwise optimization}
In a case where the dynamics law of $X(t)$ is assumed to be known, there is  a possibility to estimate $J$ using so-called pathwise optimal control.

Up to the end of this section, we assume that
$\{\F_t\}$  is the filtration  generated by a Wiener process $W(t)$ taking the values in $\R^n$.
We assume that $X(t)$ is  an RCLL stochastic  process
adapted to $\F_t$.

Consider  a linear normed space $\oo\X=L_2([0,T];L_2(\O,\F_T,\P;\R^n))$. Let $\X$ be the closed subspace
obtained as the closure of the set of all progressively measurable with respect to $\{\F_t\}$
processes from $\X$.

 For $y\in\G$, let $\oo U(t,y)\subset \oo \X$ be the set of processes $u(s)=(u_1(t),...,u_n(t)):[t,T]\times \O\to K$
being  $\F_T$-measurable for all $t$  and such that $y+\int_t^T   u(s)ds\in \G$ a.s.

\par
Assume that we are given $y_0\in \R^m$.
 Let us consider an optimal control problem
\baa
\hbox{Maximize}\quad \E F(u,0)\quad\hbox{over}\quad u\in U(y_0,0).
\label{p}
\eaa
Let $y_u^{(0)}=u$, $y_u^{k}(t)=\int_0^ty_u^{(k-1)}(s)ds$, $k=1,2,3,...$,
where $y_u(t)=\int_0^tu(s)ds$.

Consider  a linear normed space $\oo V=L_2([0,T];L_2(\O,\F_T,\P;\R^{n\times n}))$. Let $V$ be the closure of the set of all progressively measurable with respect to $\{\F_t\}$
processes from $\oo V$.
\begin{lemma}\label{lemmaA} \index{[Dokuchaev (2015c)]}  Let $u\in \oo U(0,y_0)$. The following statements are equivalent:
\begin{enumerate}
\item
$u\in U(0,y_0)$.
\item
There exists $k>1$ such that $y_u^{(k)}\in \X$.
\item
For any $k\ge 0$, $y_u^{(k)}\in \X$.
 \item
 For any $k\ge 0$ and any $v\in V$,
 \baaa
 \E\int_0^T \mu(t)^\top y_u^{(k)}(t) dt=\E\int_0^T \mu^{(k)}(t)^\top u(t)dt=0.
 \eaaa
 Here $M(t)=\int_0^t v(s)dW(s)$, $\mu(t)=M(T)-M(t)$,
 $\mu^{(0)}=\mu$, $\mu^{(k)}(t)=-\int_t^T\mu^{(k-1)}(s)ds$, $k=1,2,3,...$.
\end{enumerate}
\end{lemma}

Let $k\in\{0,1,2,..\}$ be selected.

For $v\in V$, let $M(t)=\int_0^t v(s)dW(s)$. Clearly,  $M(T)\in
 L_2(\O,\F_T,\P)$ and $M(t)=\E_t M(T)$. Set $\mu(t)=M(T)-M(t)$. For
 $u\in\oo U(0,y_0)$, $v\in V$, and $\mu^{(k)}=\mu^{(k)}(\cdot,v)$ defined as above, introduce Lagrangian
\baaa \L(u,v)=F(u,0)+\E\int_0^T \mu^{(k)}(t)^\top u(t)dt. \eaaa
The following theorem  is similar to Theorem 5.1 from  Dokuchaev (2015a). A related result is presented in Theorem 5.1 in Bender and Dokuchaev (2016b).
\begin{theorem}\label{ThD}
\baa
 \sup_{u\in
U(0,y_0)} \E F(u,0) =\sup_{u\in\oo U(0,y_0)}\inf_{v\in V} \L(u,v)= \inf_{v\in V} \sup_{u\in\oo U(0,y_0)} \L(u,v).
\label{infsup} \eaa
\end{theorem}

It can be noted that Theorem \ref{ThD}  does not establish the existence
of a saddle point. However, it can be used to estimate the value $\sup_{u\in\oo U}\E F(u,0)$ using  Monte-Carlo simulation of
$\mu^{(k)}$ and pathwise solution of the problem $\sup_{u\in\oo U} \L(u,v)$ in the spirit of the methods developed in
Davis and Burstein (1992), Rogers (2007), Andersen and Broadie (2004), Haugh and Kogan (2004),  Bender (2011), Brown et al. (2010), Bender and Dokuchaev (2016a,b);
this supremum can be found using pathwise optimization in the class of {\em anticipating} controls $u\in \oo U$ that do not have to be adapted. An advantage of this approach is that it seeks only the
solution starting from a particular $y(0)$, whereas the HJB approach described above
requires calculating the solution from all starting points.
The papers  mentioned here suggest running Monte-Carlo over a set of martingales that are considered to be
independent variables for the Lagrangian. In the term of Theorem \ref{ThD}, this means
maximization over the set $v\in V$.  Unfortunately, this set is quite wide. On
the other hand, the optimal martingale has a very particular
dependence on the underlying stochastic process and optimal value function, in the cases of some known explicit solutions. For  example, for a related problem
considered in  Bender and Dokuchaev (2016b), the corresponding optimal martingale was found to be $\mu(t)=D_y J(t,Y(t))$, where
$Y(t)$ was an optimal state process $J(t,y)$ was the optimal value function for the problem satisfying a First order BSPDE
being an analogue of the HJB  equation  (Theorem 5.1 in Bender and Dokuchaev (2016b)).
 This shows that a sequence of randomly generated  martingales may not attend a close proximity of
 the optimal  martingale in a reasonable time. Theorem \ref{ThD}
  allows one to replace simulation of martingales by simulation of  more special
processes $\mu^{(k)}(t)$ that are $k-1$ times pathwise differentiable, with
absolutely continuous derivative  $d^{k-1}\mu^{(k)}(t)/dt^{k-1}$. After time discretisation, these processes
can be presented  as  processes with a reduced range of finite differences of order $k$. This could help to reduce the calculation time.
The case where $k=0$ corresponds to martingale duality was studied  in
Davis and Burstein (1992), Rogers (2002), Andersen and Broadie (2004), Haugh and Kogan (2004),  Bender (2011b), Brown et al. (2010), Bender and Dokuchaev (2016a,b).

\section{The proofs}  \index{(The case where $G=0$, $m=n=1$)}

{\em Proof of Lemma \ref{lemmaE}.} For the case where $G\neq 0$, the proof follows from  the standard properties of the quadratic forms. The proof for $G=0$
repeats the proof from Bender and Dokuchaev (2016a).
\comm{
Let us assume that $G=0$.
 We consider $U(t,Y)$ as a subset of $L^2(\mathcal{F}\otimes \mathcal{B}[0,T], P \otimes \lambda_{[0,T]})$ where $\mathcal{B}[0,T]$ and
$\lambda_{[0,T]}$ denote the Borel $\sigma$-field and  the Lebesgue measure on $[0,T]$. Note that $U(t,Y)$ is a weakly sequentially compact
subset
of the Hilbert space $L^2(\mathcal{F}\otimes \mathcal{B}[0,T], P \otimes \lambda_{[0,T]})$, because $U(t,Y)$ is bounded and closed in the strong topology and convex.
\par
We now introduce the set
$$
\mathcal{M}=\left\{ E[\int_t^T u(r)X(r)dr|\mathcal{F}_t],\; u\in U(t,Y) \right\}.
$$
It is straightforward to check that $\mathcal{M}$ is closed under
 maximization, i.e. $M_1, M_2 \in \mathcal{M}$
implies that $M_1 \vee M_2 \in \mathcal{M}$. Hence, by Theorem A.3
in citet{KS1998}, there is a sequence $(u_n)\subset U(t, Y)$
such that
\begin{equation}\label{eq:hilf1}
E[\int_t^T  u_n(r)X(r)dr|\mathcal{F}_t] \uparrow J(t,Y),\quad n \rightarrow \infty
\end{equation} $\P$-almost surely. As $U(t,Y)$ is weakly sequentially compact, we can assume without loss of generality (by passing to a subsequence, if necessary), that $u_n$ converges
weakly in $L^2(\mathcal{F}\otimes \mathcal{B}[0,T], \P \otimes \lambda_{[0,T]})$ to some $\bar u \in U(t,Y)$. We now show that $\bar u$ is indeed
optimal. Suppose $A\in \mathcal{F}_t$.  By weak convergence of $(u_n)$ to $\bar u$ and considering $X  {\bf 1}_{A\times [t,T]}$ as an element of
$L^2(\mathcal{F}\otimes \mathcal{B}[0,T], P \otimes \lambda_{[0,T]})$, we get
\begin{equation*}
E[{\bf 1}_A E[\int_t^T u_n(r)X(r)dr|\mathcal{F}_t]] \uparrow E[{\bf 1}_A E[\int_t^T \bar u(r)X(r)dr|\mathcal{F}_t]],\quad n\rightarrow \infty,
\end{equation*}
which, combined with (\ref{eq:hilf1}), yields
$$
 \E[{\bf 1}_A E[\int_t^T \bar u(r)X(r)dr|\mathcal{F}_t]]= E[{\bf 1}_A J(t,Y)]
$$
As $A\in \mathcal{F}_t$ was arbitrary, this immediately gives
$$
J(t,Y) = E_t[\int_t^T  \bar u(r)X(r)dr|\mathcal{F}_t].
$$
This completes the proof.
\par
On the unconditional probability space $(\O,\P,\F)$, we consider the
Hilbert space $\H_t$ of all square integrable and adapted processes
from $L_2([t,T]\times\O,\P(\cdot|\F_t),\F)$, with the "conditional"
norm $\|u\|_{\H_t}=\left(\E_t\int_0^T|u(s)|^2ds\right)^{1/2}$.

Let $\tau$ be a Markov time with values at $[t,T]$, and let $Y$ be
$\F_t$-measurable random variable.

We have that $U(t,Y)$ is bounded and  closed subset of $\H_0$; it
is convex and hence weakly closed. Therefore, $U(t,Y)$ is weakly
compact in $\H_0$.

Let $\H=\{\E_t\int_t^T u(t)X(t)dt,\quad u\in U(t,Y)\}.$ The
set $\H$ is max-stable, meaning that if $h_1,h_2\in\H$ then
$\max(h_1,h_2)\in \H$. It can be seen from the fact that if
$h_i=\E_t\int_t^T u_i(t)^\top X(t)dt$ then
$h=\max(h_1,h_2)=\E_t\int_t^T u(t)^\top X(t)dt,$ where
$u=\Ind_{\{t>t\}}(\Ind_{\{h_1\ge
h_2\}}u_1+\Ind_{\{h_2>h_1\}}u_2)$.

By Theorem A.3 [KS2], it follows that there exists a sequence
$\{u_n\}$ such that the sequence $h_n=\E_t\int_t^T
u_n(t)^\top X(t)dt$ is such that $\esssup_{\o} h_n$ is non-decreasing
monotonically.

Since $U(t,Y)$ is weakly compact in $\H_0$, there exists a
subsequence $\{u_m\}$ with the weak limit $u_*$. Let
$h_*=\E_t\int_t^T u_*(t)X(t)dt$.

Let $A=\cup_m\{\o:\ h_m(\o)>h_*(\o)\}$. In this case, $\E \Ind_A
h_m$ is non-decreasing in $m$ and \baaa \E \Ind_A
h_m=\E\Ind_0^T\Ind_{\{t\in[t,T]\}}\Ind_A u_m(t)^\top X(t)dt \to \E
\Ind_A h_* \E\Ind_0^T\Ind_{\{t\in[t,T]\}}\Ind_A
u_*(t)^\top X(t)dt\quad\hbox{as}\quad m\to +\infty.\eaaa This means that
\baa\E \Ind_A h_m \le \E \Ind_A h_*.\label{ner1}\eaa Suppose that
$\P(A)>0$, i.e., there exists $m_0$ such that $\P(A_m)>0$ for all
$m>m_0$. We have that \baaa\E \Ind_{A} (h_m-h_*)=\E \Ind_{A_m}
(h_m-h_*)+\E \Ind_{A\backslash A_m} (h_m-h_*).\eaaa By the
definitions, $\E \Ind_{A_m} (h_m-h_*)\ge 0$. In addition,
$\E\Ind_{A\backslash A_m} (h_m-h_*)\to 0$ as $m\to +\infty$, since
$\P(A\backslash A_m)\to 0$. This contradicts (\ref{ner1}). Hence
$\P(A)=0$, i.e., $h_*\ge h_n$ a.s.. Hence $u_*$ is an optimal
strategy. This completes the proof. $\Box$}
\par
{\em Proof of Lemma \ref{lemmaJ}.} To prove statement (i), it suffices to observe that $|u(t)^\top X(t)|\le L|X(t)|$ if $G=0$ and that optimal $u$
cannot be too large for $G\neq 0$.
Statement (ii) follows from the definition of $J$ and
from the fact that $U(t,\w y)\subset U(t,y)$.
Let us prove statement (iii). \index{NONZERO G IS ALSO COVERED}
Let $y_j<\w y_j$, $y=(y_1,...,y_n)^\top$, $\w y=(y_1,...y_{j-1},\w y_j,y_{j+1},...,y_n)^\top$. Clearly, $J(t,\w y)\le J(t,y)$ a.s. for
any $t$. In addition,  \baaa J(t,y)&\le& J(t,\w y)+\esssup_{v\in
U(t,y):\ \int_t^Tv(s)ds-(\w y-y)\in\G_0} \E_t\int_t^T |X(s)^\top v(s)|ds\\
&\le& J(t,\w y)+\const\cdot\E_t\esssup_{s\in[t,T]}|X(s)|(\w y_j-y_j) \quad
\hbox{a.s.}. \eaaa  This completes the proof  of Lemma \ref{lemmaJ}. $\Box$
\par
{\em Proof of lemma \ref{lemmaconc}.} It suffice to  show that $\frac{1}{2}
[J(y_1,t)+J(y_2,t)]\le J((y_1+y_2)/2,t)$ a.s.
 We have \baaa
&&\frac{1}{2}[J(y_2,t)+J(y_1,t)] =\sup_{u\in U(t,y_1)}\frac{1}{2}
F(X,u,t)+\sup_{u\in U(t,y_2)}\frac{1}{2}F(X,u,t)\\ &&= \sup_{u_1\in
U(t,y_1), u_2\in
U(t,y_2)}\frac{1}{2}[F(X,u_1,t)+F(X,u_2,t)]\\&&\le \sup_{u_1\in U(t,y_1),
u_2\in U(t,y_2)}F(X,(u_1+u_2)/2,t)\le \sup_{u\in
U(t,(y_1+y_2)/2)}F(X,u,t)\\&&=J((y_1+y_2)/2,t) \quad \hbox{a.s.} \eaaa
This completes the proof  of Lemma \ref{lemmaconc}. $\Box$

The remaining part of this section is devoted to the proof of Theorem \ref{ThM}.
\subsection{Some preliminary results}
 For $y\in\G$, $t\in[0,T]$, $\t\in(t,T]$, let $U(t,\t,y)$ be the set of processes $u(s)=(u_1(s),...,u_n(s)):[t,\t]\times \O\to K$
being adapted to $\F_s$ and such that $y+\int_t^\t u(s)ds\in \G$ a.s..
\begin{lemma}\label{dpp}[Dynamic programming principle] For any $y\in[0,1]$ and any
time $\t\in(t,T)$, \baaa J(t,y)=\sup_{u\in
U(t,\t,y)}\E_t\left[\int_t^\t f(X(s),u(s),t)ds+J(y+\nu_u,\t)\right]. \eaaa
\end{lemma}
\par
{\em Proof.} It suffices to prove that
 \baaa J(t,y)\le \sup_{u\in
U(t,\t,y)}\E_t\left(\int_t^\t f(X(s),u(s),s)
ds+J(y+\nu_u,\t)\right),  \eaaa where $\nu_u\defi \int_t^\t u(s)ds$. We have \baaa &&J(t,y)\le
\sup_{u\in U(t,y)}\E_t\left(\int_t^\t f(X(s),u(s),s) ds+\int_\t^T f(X(s),u(s),s) ds \right)\\
&&= \sup_{u\in U(t,\t,y)}\,\sup_{v\in U(\t,T,y+\nu_u)}\E_t\left(\int_t^\t f(X(s),u(s),s) ds+\E_\t\int_\t^T f(X(s),v(s),s)ds \right)
\\
&&= \sup_{u\in U(t,y)}\E_t\left(\int_t^\t f(X(s),u(s),s)ds+\sup_{v\in U(\t,T,y+\nu_u)}\E_\t\int_\t^T f(X(s),v(s),s)ds \right).
\eaaa  This completes the proof. $\Box$

Let $t_k=kT/N$, $N\in\{1,2,...\}$, $k=0,1,2,...,N$.

 Let   $U_N(s,y)$ be the set of all admissible $u\in U(s,N)$ such that $u(t)\equiv u(t_k)$, $t\in[t_k,t_{k+1})$, $t_k=kT/(N-1)$, $k=0,1,...,N-1$.

Let \baaa&&\V(t,y)=\Bigl\{ v\in L_1(\O,\F_{t_k},\P,\R^m),\quad\hbox{where}\quad t_k=\max\{t_m:\ t_m\le t\},
\quad\\&&\hphantom{xxxxxxxxxxxxxx}\hbox{such that}\quad v\in K \quad\hbox{a.e.},\quad
y+(t_{k+1}-t)  v\in \G\quad \hbox{a.s.}\Bigr\}.\eaaa
 \begin{lemma} \label{lemma1} Assume that $X(t)=X(t_k)$ \comm{is $\F_{t_k}$-measurable} for $t\in[t_k,t_{k+1})$, $k=0,1,...,N-1$. Then, for any $(s,y)$, there exists optimal in $U(t,y)$ process $u\in U_N(s,y)$. In addition,  \baa J(t,y)=\esssup_{v\in \V(t,y)}
\left[(t_{k+1}-t)[X(t_k)^\top v-v^\top G v]+\E_tJ(t_{k+1},y+(t_{k+1}-t)v)\right]\label{Jv1}\eaa
for $t\in[t_k,t_{k+1})$. The corresponding $v\in\V(t,y)$ exists for all $(t,y)$.
\end{lemma}

{\em Proof of  Lemma \ref{lemma1}.}  Let $k$ be given, $t\in [t_k,t_{k+1})$.  For $u\in U(t,t_k,y)$, let
$\nu_u=\int_t^{t_{k+1}}  u(s)ds$, $\oo \nu_u=\E_t\nu_u$, and $\ww \nu_u=\oo\nu_u/(t_{k+1}-t)$. We have that
\baa J(t,y)&=&\sup_{u\in U(t,t_k,y)}\E_t\left[\int_t^{t_{k+1}}f(X(s),u(s),s) ds+J(t_k,y+\nu_u)\right]
\nonumber\\&=&\sup_{u\in U(t,t_k,y)}\left[X(t_k)^\top \oo\nu_u-\E_t\int_t^{t_{k+1}}u(s)^\top G u(s)ds +\E_tJ(t_{k+1},y+\nu_u)\right] . \label{ququ}\eaa  By the concavity of $J(t,y)$ in $y$, it follows that
$\E_tJ(t_{k+1},y+\nu_u)\le \E_tJ(t_{k+1},y+\oo\nu_u)$.  By the concavity of $f(x,u,s)$ in $u$, it follows that
\baaa J(t,y)=\sup_{u\in U(t,t_k,y)}\left[X(t_k)^\top \oo\nu-(t_{k+1}-t)\ww\nu_u^\top G\ww \nu_u +\E_tJ(t_{k+1},y+\nu_u)\right] . \eaaa

 Clearly,
any $\F_{t_k}$-measurable  $u(s)|_{s\in[t,t_{k+1}]}$ is optimal for (\ref{ququ}) if
 $(t_{k+1}-t)v=\nu_u=\oo\nu_u$ for  $v$  being optimal for (\ref{Jv1}).  This completes the proof of  Lemma \ref{lemma1}. $\Box$
\subsection{The case of piecewise constant $X$}
 \begin{proposition}\label{prop0} Assume that $X(t)=X(t_k)$ \comm{is $\F_{t_k}$-measurable} for $t\in[t_k,t_{k+1})$ (i.e., the assumptions of Lemma \ref{lemma1}
hold).  Then Theorem \ref{ThM} holds.
\end{proposition}
\par
{\em Proof.}  For $k=N-1,N-2,...1,0$, let
\baaa
V_k=\{v\in L_\infty([t_k,t_{k+1}],\R^n):\quad v(t)\in K\},
\eaaa  and let functions $J_k:[t_k,t_{k+1}]\times \G\times\O\to\R$  be defined consequently as the value functions  for deterministic
(on  the conditional
probability spaces given $\F_{t_k}$) control problems
\baa
&&\hbox{Maximize}\quad \w J_{k+1}(\tau_k^{x,s},y^{x,s}(\tau_k^{x,s}))+\int_{s}^{\tau_k^{x,s}}f(X(t),v(t),t)dt\quad\hbox{over}\quad v\in V_k,\nonumber\\
&&\hbox{subject to}\quad \frac{dy^{x,s}}{dt}(t)= v(t),\quad y^{x,s}(s)=x,\nonumber\\
&&
\hphantom{\hbox{subject to}\quad}\tau_k^{x,s}=t_{k+1}\land \inf\{t>s:\ y^{x,s}(t)\notin\G\}.
\label{pk}\eaa
Here  \baaa
\w J_k(t,y)\defi\E_t J_k(t,y).
\eaaa
We assume that $\w J_{N}(T,y)\equiv 0$.

In other words,
\baaa
J_k(y,s)=\sup_{v\in V_k} \left(\w J_{k+1}(\tau_k^{x,s},y^{x,s}(\tau_k^{x,s}))+\int_{s}^{\tau_k^{x,s}}f(X(t),v(t),t)dt\right).
\eaaa
By the assumptions  that $X$ is piecewise constant,
 we have  that \baa
\F_t=\F_{t_k}, \quad t\in [t_k,t_{k+1}).
\label{Fk}\eaa
It follows that $J_k(t,\cdot)$ are $\F_{t_k}$-measurable for $t\in [t_k,t_{k+1}]$. On the conditional
probability space given $\F_{t_k}$, the values $J_{k}(t,x)$ can be deemed  to be deterministic for $t\in [t_k,t_{k+1}]$.

In addition, we have that each $J_k$ is the value function for the problem
\baaa
&&\hbox{Maximize}\quad \w J_{k+1}(t_{k+1},y^{x,s}(t_{k+1}))+\int_{s}^{t_{k+1}}f(X(t),v(t),t)dt\quad\hbox{over}\quad v\in \oo U_k(s,x),\\
&&\hbox{subject to}\quad \frac{dy^{x,s}}{dt}(t)= v(t),\quad y^{x,s}(s)=x.\eaaa
Here $x\in\G$, $s\in[t_k,t_{k+1})$,
$\w J_{k+1}(t,y)=\E_{t_k} J_{k+1}(t,y)$, \baaa
\oo U_k(s,x)=\{v\in L_\infty([t_k,t_{k+1}],\R^n):\quad v(t)\in K,\quad x+\int_s^{t_{k+1}} v(r)dr\in \G\}.\eaaa

By Lemma \ref{dpp}, we obtain, consequently for  $k=N-1,N-2,...,1,0$, that
\baa
J(t,y)=J_k(t,y), \quad t\in [t_k,t_{k+1}).
\label{JJk}\eaa

The proof below is for the special case where $n=m$, $X_k(t)\ge 0$, and either $G\neq 0$ or $K=\R^n$.

Let us show that   $J=J_{k}$ are unique viscosity solutions of the boundary value problems
\baa
&& \frac{\p J}{\p t}(t,y)+\sup_{v\in K}(f(y,v,t)+v^\top D_y^+J(t,y)), \qquad t\in[t_k,t_{k+1}],\quad y\in\G,\nonumber \\
&&J(t_{k+1},y)=\w J_{k+1}(t_{k+1},y),\quad J(t_{k+1},y)|_{y\in\p\G}=0.
\label{BSPDE3}\eaa
Possibly, this statement follows from the known theory for deterministic controlled equations with  first order HJB equations.
For this, we would need the existence of a solution for HJB equations (\ref{BSPDE3}) plus
a verification theorem connecting them with the control problem.
So far, we were  unable to find a result covering our case; the closest one is Theorem 5.2 from Garavello (2003);
 \index{V11} unfortunately, condition (iv) in this theorem does not
  hold in our case. Other existing literature covers either domains without boundaries
or  bounded domains; see e.g. Lions (1982). \index{V4} The existing results for solvability  of first order HJB equations require additional conditions such as $C^1$-smooth boundary functions; see e.g.
Bressan  and Flores (1994),  Crandall (1983),
Dacorogna  and Marcellini  (1998). \index{V1,V2,V3}
  To overcome these difficulties, we derived  below equation (\ref{BSPDE3}) for $J=J_k$  directly using the features imposed by our special setting.

Let $\w y(t,\t)=y+v(\t-t)$, $t\in[t_k,t_{k+1})$, $\t\in[t,t_{k+1})$,  for an optimal $v$ in (\ref{Jv1}). Clearly, the strategy
$u(s)=v$ is the part on $[t_k,t_{k+1})$ of the optimal solution for the problem  (\ref{J}) with $t=t_k$. Respectively, the process
$(u(s),y(s))=(v,\w y(t,s))$ is the part on $[t_k,t_{k+1})$ of the optimal process for problem  (\ref{J}).
Therefore, $v\in \V(t,y)$ is an optimal point for (\ref{Jv1}) if and only if, for any $\t\in[t,t_{k+1})$, it is a maximum point for the problem
\baa J(t,y)=\esssup_{v\in \V(t,y)}
\left[(\t-t)[v^\top X(t_k)
-v^\top Gv]+\E_tJ(\t,\w y(t,\t))\right]. \label{Jv}\eaa
 \begin{lemma}\label{lemma2G} Assume that $X(t)=X(t_k)$ for $t\in[t_k,t_{k+1})$, $k=0,1,...,N-1$.
 Assume that $G\neq 0$ and $K=\R^n$.
Then  the  following holds.
\begin{enumerate} \item The only optimal point of (\ref{Jv}) is
 $v=\frac{1}{2}G^{-1} [X(t_k)+D^+_{y}J(\t,\w y(t,\t))]$.
 \item For any $t\in[t_k,t_{k+1})$, $\t\in[t,t_{k+1})$,   \baa
J(t,y)&=&
(\t-t)[v^\top X(t_k)-v^\top G v]+J(\t,y+v(\t-t))\nonumber
\\ &=&(\t-t)[v^\top X(t_k)-v^\top G v]+J(\t,\w y(t,\t)).\label{JG}\eaa
 \item
For $t\in[t_k,t_{k+1})$, \baa
 J(t,y)&=& \frac{1}{4}\int_t^{t_{k+1}}\left(X(s)+D^+_{y}J(s,y)\right)^\top G^{-1} \left(X(s)+D^+_{y}J(s,y)\right) ds\nonumber\\&+&\w J_{k+1}(t_{k+1},y).\hphantom{xxx} \label{DJ0}\eaa
\end{enumerate}
\end{lemma}
\par
{\em Proof of Lemma \ref{lemma2G}.}  Statements (i)-(ii) follow immediately from (\ref{JG}) and
from the optimality of $v$. \index{Statement (i) follows from the necessary conditions of optimality since $-X(t)\in  [D^+_{y_k}J(t,y),D^-_{y_k}J(t,y)]$.   Statement (iii) follows from (\ref{Jv}) and
from the optimality of $v=L$.}
Let us show that \baa
 J(t,y)=v^\top \int_t^{t_{k+1}}\left(X(s)-Gv+D^+_{y}J(s,y)\right)ds +\w J_{k+1}(t_{k+1},y). \label{DJG}\eaa
By  (\ref{JG}), we have that   \baaa D_tJ(t,y)&=&
-X(t_k)^\top  v+v^\top G v-D^+_{y}J(\t,y+ v(\t-t))^\top v  \\ &=&
-X(t_k)^\top  v+v^\top G v-D^+_{y}J(\t,\w
y(t,\t))^\top v. \eaaa
It follows that $D^+_{y}J(\t_1,\w
y(t,\t_1))^\top v=D^+_{y}J(\t_2,\w
y(t,\t_2))^\top v$ for any $\t_1,\t_2\in[t_k,t_{k+1}]$.
If we assume some additional regularity of $D_y^+J$, we have that  $D^+_yJ(t,
y)^\top v=D^+_yJ(\t,\w
y(t,\t))^\top v$ and, therefore, \baaa D_tJ(t,y)&=&-v^\top (X(t)+D^+_{y}J(t,
y))+v^\top Gv\\&=&-\frac{1}{4}\left(X(t)+D^+_{y}J(t,y)\right)^\top G^{-1} \left(X(t)+D^+_{y}J(t,y)\right) . \eaaa  This gives (\ref{DJ0}).
Without these additional regularity assumptions, we prove (\ref{DJ0}) as the following.   We observe that
\baaa J(t,y)=
v^\top\int_t^{t_{k+1}}\left(X(s)-Gv+D^+_yJ(t_{k+1},\w y(s,t_{k+1}))\right)ds+\w J_{k+1}(t_{k+1},y)\\
 = v^\top\int_t^{t_{k+1}}\left(X(s)-Gv+D^+_yJ(s+\e,\w y(s,s+\e))\right)ds+\w J_{k+1}(t_{k+1},y)
 \eaaa for any $\e\in (0,t_{k+1}-t_k)$.  Hence
 \baaa
 J_k(t,y)= \frac{1}{4}\int_t^{t_{k+1}}\left(X(s)+D^+_{y}J(s,y)\right)^\top G^{-1} \left(X(s)+D^+_{y}J(s,y)\right) ds
 +\w J_{k+1}(t_{k+1},y)\\+\xi_\e(t,y), \eaaa
where
\baaa
\xi_\e(t,y)=\frac{1}{4}\int_t^{t_{k+1}}\biggl[
\left(X(s)+D^+_{y}J(s+\e,\w y(s,s+\e))\right)^\top G^{-1} \left(X(s)+D^+_{y}J(s+\e,\w y(s,s+\e))\right)\\
-\left(X(s)+D^+_{y}J(s,y)\right)^\top G^{-1} \left(X(s)+D^+_{y}J(s,y)\right)\biggr]
   ds.
\eaaa
For an arbitrarily selected $\eta\in L_\infty(0,1)$, let $\g_\e(t)=\int_0^1 \xi_\e(t,y)\eta(y)dy$. By Luzin's theorem, we have
$\g_\e(t)\to 0$ as $\e\to 0$. Hence (\ref{DJG}) holds. Substitution of optimal $v$ gives the desired formula.
This completes the proof of Lemma \ref{lemma2G}. $\Box$.

Let \baaa
 &&\wD_{y_i}J(t,y)=h\in [D^+_{y_i}J(t,y),D^-_{y_i}J(t,y)]
 \\&& \hbox{such that}\quad |h+X_i(t)|=\min_{g\in [D^+_{y_i}J(t,y),D^-_{y_i}J(t,y)] } |g+X_i(t)|.\eaaa
Clearly, $h$ in this definiion is unique.
\begin{lemma}\label{lemma2}
Assume that $X(t)=X(t_k)$ for $t\in[t_k,t_{k+1})$, $k=0,1,...,N-1$. Assume that $K=[0,L]^n$ and $G=0$.
Let $t\in[t_k,t_{k+1})$ and $y\in\G$ be given, and  let $v\in\V(t,y)$ be an optimal point  in (\ref{Jv1}) and
 hence optimal in
 (\ref{Jv}) for all $\t\in [t,t_{k+1})$), where  $\w y(t,\t)=y+v(\t-t)$, $t\in[t_k,t_{k+1})$, $\t\in[t,t_{k+1})$.

\index{NE NADO: Let $t_{k+1}\le \tau_{k}$, where $\tau_k=\inf\{s: \w y(t_k,s) \notin\G\}$}

Then,  under the assumptions of Lemma \ref{lemma1},  the following holds.
\begin{enumerate} \item  $J(t,y)=
v^\top X(t_k)(\t-t)+J(\t,\w y(t,\t))$ for $t\in[t_k,t_{k+1})$ and $\t\in[t,t_{k+1})$.
\item   If $v_i=0$, then
 $X_i(t_k)+D^+_{y_i}J(\t,\w y(t,\t))\le  0$.\\
 If $v_i=L$, then
$X_i(t_k)+D^+_{y_i}J(\t,\w y(t,\t))\ge 0$.
\item If  $v_i\in (0,L)$ then
 $X_i(t_k)+\w D_{y_i}J(\t,y(t,\t))=0$.
\item
For $t\in[t_k,t_{k+1})$, \baa
 J(t,y)=\sum_{i=1}^nL \int_t^{t_{k+1}}\left(X_i(s)+\wD_{y_i}J(s,y)\right)_+ds +\w J_{k+1}(t_{k+1},y). \label{DJ}\eaa
\end{enumerate}
\end{lemma}
\par
{\em Proof of Lemma \ref{lemma2}.}  Statement (i) follows immediately from  (\ref{Jv}) applied
with the optimal $v$. Statement (ii) follows from (i).
Consider optimization on the conditional probability space given $\F_{t_k}$.
Let us prove statement (iii).
If $v_i\in (0,L)$ is optimal for (\ref{Jv}), then it is optimal for $t_{k+1}$ replaced by any $\t\in(t_k,t_{k+1}]$, then  \baaa
0\in [ X_i(t_k)+D_{y_i}^+J(t,\w y(t,\t)),X_i(t_k)+D_{y_i}^-J(t,\w y(t,\t))].\eaaa Hence  \baa
X_i(t_k)+\w D_{y_i} J(t,\w y(t,\t))=0 .\label{nil}\eaa

Let us prove statement (iv).
By the definition of $\wD$ and by concavity of $J$ in $y_i$, we obtain the following.
\begin{itemize}
\item
if $X_i(t)+D^+_{y_i}J(t,
\w y(t,\t))\ge 0$, then $X_i(t)+\wD_{y_i}J(t,
\w y(t,\t))\ge 0$.
\item If $X_i(t)+D^+_{y_i}J(t,
\w y(t,\t))< 0$, then \baaa (X_i(t)+ D^+_{y_i}J(t,
\w y(t,\t)))_+=(X_i(t)+\wD_{y_i}J(t,
\w y(t,\t)))_+ = 0.\eaaa
\end{itemize}
For all cases listed in statements (ii)-(iii), we have  \baaa
v_i(X_i(t)+D^+_{y_i}J(t,
\w y(t,\t)))=L(X_i(t)+D_{y_i} ^+J(t,\w y(t,\t)))_+.
\eaaa
Therefore, for all cases listed in statements (ii)-(iii), we have  \baaa
&&v_i(X_i(t)+D^+_{y_i}J(t,
\w y(t,\t)))=L(X_i(t)+D_{y_i} ^+J(t,\w y(t,\t)))_+\\ &&=v_i(X_i(t)+\w D_{y_i}J(t,
\w y(t,\t)))=L(X_i(t)+\w D_{y_i} J(t,\w y(t,\t)))_+.
\eaaa

By  statement (i),  we have that  \baaa D_tJ(t,y)=
-X(t_k)^\top  v-D^+_{y_i}J(\t,\w
y(t,\t))^\top v. \eaaa
It follows that $D^+_{y}J(\t_1,\w
y(t,\t_1))^\top v=D^+_{y}J(\t_2,\w
y(t,\t_2))^\top v$ for any $\t_1,\t_2\in[t_k,t_{k+1}]$.
Some additional regularity of $D_y^+J$ gives that  $D^+_yJ(t,
y)^\top v=D^+_yJ(\t,\w
y(t,\t))^\top v$ and, therefore, \baaa D_tJ(t,y)=-v^\top (X(t)+D^+_{y}J(t,
y)). \eaaa
Without these additional regularity assumptions, we prove (\ref{DJ}) as the following.   We observe that
\baaa J(t,y)=v^\top\int_t^{t_{k+1}}\left(X(s)+D^+_yJ(t_{k+1},\w y(s,t_{k+1}))\right)ds+\w J_{k+1}(t_{k+1},y)\\
 = \sum_{i=1}^nL\int_t^{t_{k+1}}\left(X_i(s)+D^+_{y_i}J(s+\e,\w y(s,s+\e))\right)_+ ds+\w J_{k+1}(t_{k+1},y),
 \eaaa for any $\e\in (0,t_{k+1}-t_k)$.  Hence
 \baaa
 J_k(t,y)= \sum_{i=1}^nL\int_t^{t_{k+1}}\left(X_i(s)+D^+_{y_i}J(s,y)\right)_+ ds +\w J_{k+1}(t_{k+1},y)+\xi_\e(t,y), \eaaa
where
\baaa
\xi_\e(t,y)
=\sum_{i=1}^nL\int_t^{t_{k+1}}\biggl[\left(X_i(s)+D^+_{y_i}J(s+\e,\w y(s,s+\e))\right)_+  -
\left(X_i(s)+D^+_{y_i}J(s,y)\right)_+ \biggr] ds .
\eaaa
For an arbitrarily selected $\eta\in L_\infty(0,1)$, let $\g_\e(t)=\int_0^1 \xi_\e(t,y)\eta(y)dy$. By Luzin's theorem, we have
$\g_\e(t)\to 0$ as $\e\to 0$. Hence (\ref{DJ}) holds.
This completes the proof of Lemma \ref{lemma2}.$\Box$.
\par
\vspace{4mm}
We now in the position to prove Proposition \ref{prop0}.
By (\ref{BSPDE3}), we obtain that
\baa
&& J_k(t,y)=\w J_{k+1}(t_{k+1},y) +\int_t^{t_{k+1}} \sup_{v\in K}[f(X(s),v,s)+v^\top  D_y^+J_k(s,y)]ds,\nonumber\\ &&\hphantom{MMMMMMMMMMMMMMM} t\in[t_k,t_{k+1}],\quad y\in\G,\nonumber \\
&&J_k(t_{k+1},y)|_{y\in\G}=0.
\label{BSPDE5}\eaa
This can be rewritten as
\baa
&& J_k(t,y)=\E\left\{\left.\w J_{k+1}(t_{k+1},y) +\int_t^{t_{k+1}} \sup_{v\in K}[f(X(s),v,s)+v^\top  D_y^+J_k(s,y)]ds\right|\mathcal{F}_{t_k}\right\},\nonumber\\ &&\hphantom{MMMMMMMMMMMMMMM} t\in[t_k,t_{k+1}],\quad y\in\G,\nonumber \\
&&J_k(t_{k+1},y)|_{y\in\G}=0.
\label{BSPDE55}\eaa
It follows from (\ref{JJk}) that (\ref{BSPDE}) holds for $J$. This completes the proof of
Proposition \ref{prop0}.  $\Box$.

\begin{proposition}\label{limJ}
For $N=1,2,...,$ consider  piecewise constant processes defined as \baaa
&&X_N(t)=\frac{1}{t_{k+1}-t_k}\E_{t_k}\int_{ t_k}^{t_{k+1}}X(s)ds, \quad t\in[t_k,t_{k+1}),\\
 &&t_k=T(k-1)/(N-1),\quad k=0,1,2,...,N.\eaaa Let $J_N$ be the corresponding functions $J$ obtained for $X$ replaced by $X_N$.
 Then
\baa \E\int_0^T|X_N(t)-X(t)|^2dt\to 0\quad\hbox{as}\quad N\to +\infty,\quad \label{XNX}\\
J_N(y,t)\le J(y,t)\quad\hbox{a.s for all }t,\quad J_N\to J\quad\hbox{a.e. as}\quad N\to +\infty,\quad \label{JNJ}\\
D^+_yJ_N\to D^+_yJ\quad\hbox{a.e. as}\quad N\to +\infty.\quad \label{DJNDJ}\eaa
\end{proposition}
\par
{\em Proof.} The limit in (\ref{XNX}) follows from the properties of conditional expectations applied to the averaging on
 $[0,T]\times \O$; see e.g. Fetter (1977).

Let us prove  (\ref{JNJ}).
If $G=0$, then it follow immediately from the definitions
that
\baa
\esssup_{u\in U(t,y)}\E_t F(X_N,u,t) =\esssup_{u\in U_N(t,y)}\E_t F(X_N,u,t)=\esssup_{u\in U_N(t,y)}\E_t F(X,u,t)\nonumber\\
\to \esssup_{u\in U(t,y)}\E_t F(X,u,t)\nonumber\\ \hbox{as}\quad N\to +\infty\quad\hbox{a.s for all }\quad.
\label{FFF}\eaa
  If $G\neq 0$, then it suffices to observe that
\baaa
&&\E_t\int_t^T u(s)^\top X_N(s)ds=\E_t\int_t^T \oo u(s)^\top X_N(s)ds\eaaa and \baaa  -\int_t^Tu(s)^\top G u(s)ds\le -\int_t^T\oo u_N(s)^\top G\oo u_N(s)ds \quad\hbox{a.s for all }t,
\eaaa
where  \baaa
\oo u_N(t)=\frac{1}{t_{k+1}-t_k}\E_{t_k}\int_{t_k}^{t_{k+1}}u(s)ds, \quad s\in[ t_k,t_{k+1}).
\eaaa
\par Let us prove (\ref{DJNDJ}). We will be using the following lemma.
\begin{lemma}\label{lemmaf} For $\a\in\R$ and $\b\in(\a,+\infty)$, let $f:[\a,\b]\to\R$ and $f_N:[\a,\b]\to\R$ be  bounded, concave,
and uniformly  Lipschitz  functions \index{from $W^1_\infty(\a_1,\a_2)$} such that  $f_N(s)\to f(s)$ as $N\to +\infty$,
$f_N(s)\le f(s)$  for all $s\in [\a,\b]$. Then $D^+f_N(s)\to D^+ f(s)$ for a.e. $s$ as $N\to +\infty$.
\end{lemma}

\def\D2{\w\Delta^{(2)}_h}

{\em Proof of Lemma \ref{lemmaf}.} For $s\in (\a,\b)$ and $h>0$ such that $s+h\in (\a,\b]$, let $\D2 f\defi f(s+h)-f(s)-D^+f(x)h$.
By Kachurovskii's theorem, we have that $\D2 f\le 0$ and  $\D2 f_N\le 0$  for all $h$ and $N$. Let $g_N\defi f-f_N$.
We have that $f=f_N+g_N$ and  $\D2 f=\D2 f_N+\D2 g_N$.
Hence   \baa
|(\D2 g_N)_-|\le |\D2 f|\quad \hbox{for all}\quad  s\in (\a,\b), h, N,
\label{DDf}\eaa
 where $(x)_-=\min(x,0)$.

 Suppose that the statement of the lemma is incorrect.
 In this case, there exists an interval $(\oo\a,\oo\b)\subset [\a,\b]$, $\oo\a<\oo\b$, and a mapping $r:(\oo\a,\oo\b)\to (0,+\infty)$
such that  $r(s)>0$ and $|D^+g_N(s)|\ge r(s)$ for all $s\in (\oo\a,\oo\b)$.
 Clearly, $D^+g_N\to 0$ weakly in $L_2(\oo\a,\oo\b)$ as $N\to +\infty$. It follows that $\mu_N\to 0$ as $k\to +\infty$, where
 \baaa\mu_N\defi \max\{|\b_1-\a_1|;\ \oo\a\le \a_1<\b_1\le \oo\b,\ \sign D^+g_N(s)=\const,\
 s\in (\a_1,\b_1)\}.\eaaa
This increasing oscillations of $g_N$  implies that (\ref{DDf}) does  not hold.  This implies (\ref{DJNDJ}) and completes the proof of Lemma \ref{lemmaf}. $\Box$
\par
 We have that, for any given $y=(y_1,..,y_n)$ and any $j\in\{1,...,n\}$, the paths $f(y_j)=J(y,t)$ and $f_N(y_j)=J_N(y,t)$
 satisfy the assumptions  of Lemma \ref{lemmaf}, for a given
 $(y_1,...,y_{j-1},y_{j+1},...,y_n)$, for  $y_j\in[\a,\b]$ and a interval $[\a,\b]$ such that $y=(y_1,...,y_{j-1},y_j,y_{j+1},...,y_n)\in\G$.
This completes the proof of Proposition \ref{limJ}.  $\Box$

We are now in the position to complete the proof of Theorem \ref{ThM}.

{\em Proof of Theorem \ref{ThM}.} Let $X_N=(X_{N1},...,X_{Nn})^\top$ and $J_N$  be such as described in Proposition \ref{limJ}.
Starting from now, we will consider a subsequence $N=N_k$, $k=1,2,...,$ such that $X_N\to X$ a.e..
It can be noted that, for any $y$, there is an integrable process $\xi(t,\o)$ such that $|X_N(t)|+|D_{y} J_N(t,y)|\le \xi(t,\o)$.

Assume that $K=[0,L]^n$, \xxxonly{$n=m$,} and $f(t,x,u)=u^\top X$. By (\ref{DJNDJ}), we have for $t<T$ and a.e. $y\in\G$ that
\baaa
\E_t\int_t^T
\left(\sum_{i=1}^n(X_{Ni}(s)+ \myD_{y_i} J_N(s,y))_+\right)ds
\to \E_t\int_t^T \left(\sum_{i=1}^n(X_{i}(s)+ \myD_{y_i} J(s,y))_+\right)ds\\\quad\hbox{a.e. as}\quad N\to +\infty.
\label{limDJDJ}\eaaa
By (\ref{JNJ}), we obtain the theorem statement for this case. The proof for the case where $G\neq 0$ is similar. $\Box$

\par
{\em Proof of Lemma \ref{lemmaA}} is straightforward and will be omitted here.
It is based on the fact that, by the martingale representation theorem
$\mu(t)=\int_t^T \w\mu(s)dw(s)$, for some $\w\mu\in \V$.
 $\Box$
\par
{\em Proof of Theorem \ref{ThD}.} The first equality  follows from Lemma \ref{lemmaA}.
Furthermore, we have that $\L(u,v)$ is concave in $u\in U$ and
 affine in $v\in V$.
 In addition, $\L(u,v)$ is continuous in $u\in L_{2}([0,T]\times\O)$ given $v\in V$, and
 $\L(u,v)$
 is continuous in $v\in V$ given $u\in U$.
 The statement of the theorem  follows Proposition 2.3 from Ekland and Temam  (1999),  Chapter VI. \index{p.175 in Russian edition}
 Statement (ii) follows from (i) and Proposition 1.2  from \index{Ekland} \index{p.172 in Russian edition}
Ekland and Temam  (1999), Chapter VI. $\Box$
\section{Discussion and future development}
The paper considers stochastic control problems in a setting  without specifying the evolution law on underlying input
 processes. The First order BSPDEs representing analogue of HJB equations are  derived; they represent further development of the concept
 originally suggested in Bender and Dokuchaev (2016a,b).  The paper extends this approach  on the  multi-dimensional state space, with different proofs,
and  establishes existence of solutions for these equations and the fact the value functions for underlying control problems
satisfy these equations.

\xxxonly{ It appears that First order BSPDEs of this type can be applied to many other
problems.

Consider, for example, the following equation \baa
&&v(t,y,z)=\E\left\{\left.\xi(y,z)+L\int_t^T(X(s)v'_z(s,y,z)+v'_y(s,y,z))_+ds\right|\F_t\right\},\quad
t\in[0,T]. \label{2D}\eaa

According to our preliminary analysis, this new equation can be used for
a variety of mathematical finance problems.

First, it can be  used for pricing of
Asian options on
{\em the volume-weighted average price (VWAP)}, with non-constant and random weight $W(t)$ defined by
the holder's consumption rate $C(t)$.
The corresponding pricing formula
requires
 to calculate
$\E\left(\int_0^{T}W(t)X(t)dt-K\right)^+$, where
\baaa
&&W(t)=C(t)\left(\int_0^TC(s)ds\right)^{-1},
 \eaaa
 for given  random processes $C(t)\ge 0$  and $X(t)$, where  $X(t)$ is the
 underlying price process. This  pricing problem is quite cumbersome  for typical classes of the  distributions of $(C,X)$.\index{see, e.g., Novikov {\em et al}
(2013).} Moreover, it is difficult to justify a particular hypothesis
about the evolution of $C$. To overcome this,  Dokuchaev (2013)
suggested calculating the price as
$\sup_{C(\cdot)}\E\left(\int_0^{T}W(t)X(t)dt-K\right)^+$. This leads
to the optimal control problem
  \baaa &&\hbox{Maximize}\quad
\E\left(\int_t^{T}W(s)X(s)ds-K\right)^+\quad\hbox{over}\quad
C(\cdot)\label{2d}
\\&&\hbox{where}\quad W(t)=C(t)\left(\int_t^TC(s)ds\right)^{-1}, \quad C(s)\in[0,L].
 \eaaa
Our conjecture that the price process satisfies
equation (\ref{2D}) with boundary conditions $ v(t,1,z)=0,\quad
v(T,y,z)=\xi(y,z)=(z/y
-K)^+.$ The corresponding state processes are
$y(t)=y+\int_s^tC(q)dq$ and $z(t)=z+\int_s^tX(q)c(q)dq$.

Further, examples of application of equation (\ref{2D}) can be found in
portfolio selection problems.  Consider a problem
that can be called a {\em problem of  optimal gradual liquidation}
of a stock portfolio. Let us consider a market with a single risky
stock with the price $X(t)$ and with a risk-free bond or money account with zero interest
rate. Let $\g(t)$ be the quantity of shares in the portfolio.  We
look for a process $\g(t)$ such that \baaa \g(0)=1,\quad
\g(T)=0,\quad \frac{d\g}{dt}(t)\in[-L,0]. \eaaa The last condition
means  that the liquidation is gradual, with a limited rate of changes
for the portfolio.  It can be reasonable, for instance, for tax
consideration. In addition, we want to gain some additional benefits
from the trading. This can be formulated as the following optimal
control problem  \baaa \hbox{Maximize}\quad \E
U\left(\int_0^T\g(t)dX(t)\right)\quad \hbox{over}\quad \g(\cdot),
\eaaa
where $U$ is a given monotonically nondecreasing utility function.
Our conjecture is  that the optimal value
for this problem is defined by equation (\ref{2D}) with the boundary
conditions $ v(t,1,z)=0,\quad v(T,y,z)=\xi(y,z)=U(z-X(0))$.

Nonlinear equations (\ref{BSPDE}) and (\ref{2D}) aroused from
optimization problems and represent versions of
Hamilton-Jacobi-Bellman equation. Examples of non-trivial linear
First Order BSDPEs can also be given. In particular, for a classical
Asian option with a constant weight, equation (\ref{2D}) has to be
replaced by a linear version  in one dimensional
space domain. For example, we have a conjecture that the price
$\E\left(\int_0^{1}X(t)dt-K\right)_+$ of an Asian option
 for a given  random process $X(t)$ of a very general kind is $v(0,0)$, where $v$ satisfies the
 following linear 1st order  equation
\baa
&&v(t,z)=\E\left\{(z-K)_++\int_t^1X(s)v'_z(s,z)ds\biggl|\F_t\right\},\quad
t\in[0,1]. \label{asian1}\eaa
}

This open ways to extend this theory on quite general models. For example, a
 possible extension is for  the case where the constraints on $\int_t^T u(s)ds$ are replaced by
 constraints on solutions of more  general  equations with input $u$. We leave it for the future research.

\subsection*{Acknowledgment}
This work  was supported by ARC grant of Australia DP120100928 to the author.

\end{document}